\documentclass[apj,numberedappendix,revtex4,twocolappendix]{emulateapj}

 \usepackage{natbib}
\usepackage{amsmath}

\newcommand{\ie}{\textit{i.e.}~}
\newcommand{\eg}{\textit{e.g.},~}
\newcommand{\SDO}{\textit{SDO}}
\newcommand{\STA}{\textit{STEREO-A}}
\newcommand{\Hinode}{\textit{Hinode}}
\newcommand{\GOES}{\textit{GOES}}

\usepackage[breaklinks,colorlinks,urlcolor=blue,citecolor=blue,linkcolor=blue]{hyperref} 
\hypersetup{
    colorlinks=true,       
     linkcolor=blue,          
     citecolor=blue,        
     filecolor=black,      
      urlcolor=blue           
}

\shorttitle{Evolution of a Current Sheet in a Solar Flare}
\shortauthors{Zhu et al.}

\begin{document}
 
\ifx \doiurl    \undefined \def \doiurl#1{\href{http://dx.doi.org/#1}{\textsf{#1}}}\fi
\ifx \adsurl    \undefined \def \adsurl#1{\href{http://adsabs.harvard.edu/abs/#1}{\textsf{#1}}}\fi
\ifx \arxivurl  \undefined \def \arxivurl#1{\href{http://arxiv.org/abs/#1}{\textsf{#1}}}\fi

\title{Observation of the Evolution of a Current Sheet in a Solar Flare}

\author{Chunming Zhu\altaffilmark{1}, Rui Liu\altaffilmark{2,3}, David Alexander\altaffilmark{4}, R.T.~James McAteer\altaffilmark{1}}
\affil{$^{1}$Department of Astronomy, New Mexico State University, NM 88003, USA; \href{mailto:czhu@nmsu.edu}{czhu@nmsu.edu}}
\affil{$^{2}$CAS Key Laboratory of Geospace Environment,
               Department of Geophysics and Planetary Sciences, 
               University of Science and Technology of China,
               Hefei 230026, China}
\affil{$^{3}$Collaborative Innovation Center of Astronautical Science and Technology, China}
\affil{$^{4}$Department of Physics and Astronomy, Rice University, TX 77005, USA}

\begin{abstract}
We report multi-wavelength and multi-viewpoint observations of a solar eruptive event which involves loop-loop interactions.  During a C2.0 flare, motions associated with inflowing and outflowing plasma provide evidence for ongoing magnetic reconnection. The flare loop top and a rising ``concave-up'' feature are connected by a current-sheet-like structure (CSLS).   The physical properties (thickness, length, temperature, and density) of the CSLS are evaluated. In regions adjacent to the CSLS, the EUV emission (characteristic temperature at 1.6 MK) begins to increase more than ten minutes prior to the onset of the flare, and steeply decreases during the decay phase. The reduction of the emission resembles that expected from coronal dimming. The  dynamics of this event imply a magnetic reconnection rate in the range 0.01 -- 0.05. 

\end{abstract}


\section{\uppercase{Introduction}}
Magnetic reconnection, a physical process involving the topological reconfiguration of magnetic fields, is generally accepted as a key mechanism for the release of free magnetic energy during solar eruptive events. Many observational features of magnetic reconnection in flares have been reported, including cusp-shaped flare loops (\citealp{tsuneta1992observation}), above-the-loop-top hard X-ray sources (\citealp{Masuda1994}), plasma inflows toward the reconnection site (\eg \citealp{Yokoyama2001a,liu2010reconnecting,su2013imaging}), plasma outflows (\eg \citealp{wang2007}; \citealp{tian2014imaging}) and their associated plasma blob ejections (\citealp{takasao2012simultaneous}) and loop shrinkage, \ie retraction of newly reconnected field lines from the reconnection site ( \citealp{vsvestka1987multi,McKenzie1999,savage2011quantitative,liu2013plasmoid}). 

During magnetic reconnection, current sheets are expected to form at the neutral region of the converging anti-parallel magnetic fields.  Several observations of CSLS were reported from X-ray and EUV images of solar flares (\eg \citealp{Sui2003,savage2010reconnection,liu2010reconnecting}) and white light coronagraph observations of coronal mass ejections (CMEs) (\citealp{Ko2003, webb2003observational, lin2007features,ciaravella2008current}). However, the physical properties and the evolution of the current sheet during a solar flare still remain unclear (see the recent review by \citealp{lin2015review}). Further, although the standard 2D flare model (\eg \citealp{Kopp1976})  successfully explains several observed features during flares, the solar flare is intrinsically a 3D phenomenon (\citealp{su2013imaging,sun2015extreme}). In this study, we present the evolution of a CSLS in a solar flare observed on 2013 December 10 in multiple  wavelengths from multiple viewpoints. The physical properties of the flaring structures and their associated dynamics are quantified. From these dynamics, we estimate the magnetic reconnection rate for this event. We also discuss the possible implications of these structures for our understanding of loop interactions, and associated reconnection, in solar flares.

\section{\uppercase{Observations}}
\subsection{Instruments}
\begin{figure*}[ht]
  \begin{center}
      \includegraphics[bb =  410 91 722 724,angle=90, clip, width=0.8\textwidth]{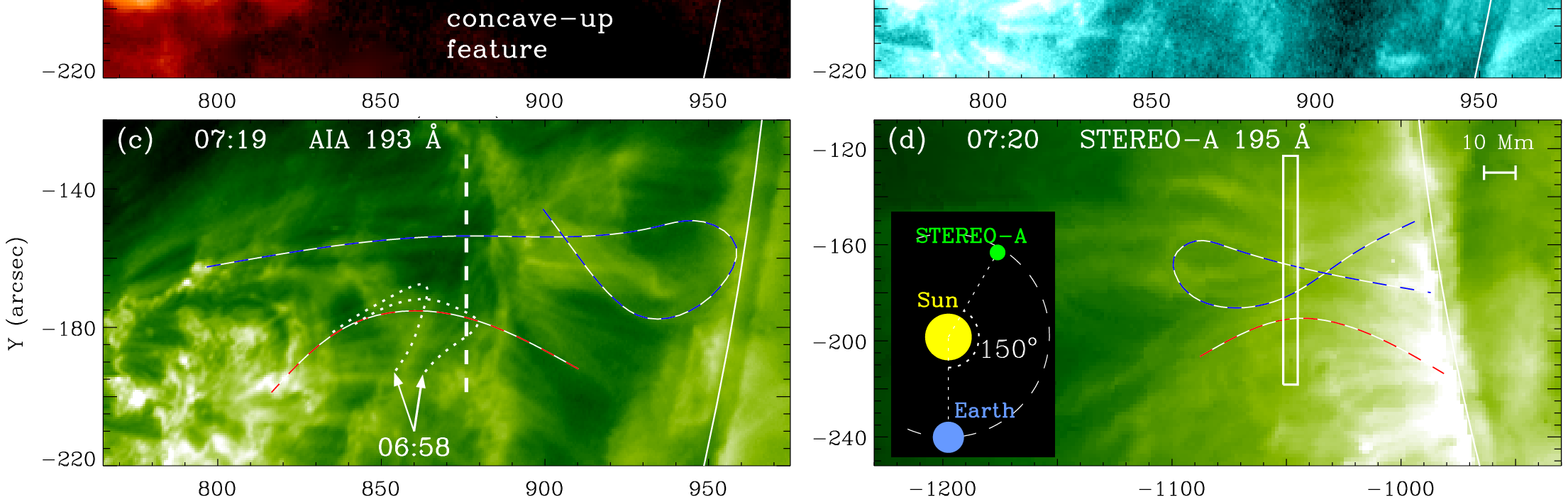}
      \includegraphics[bb =  360 253 621 730, angle=90,clip, width=0.4\textwidth]{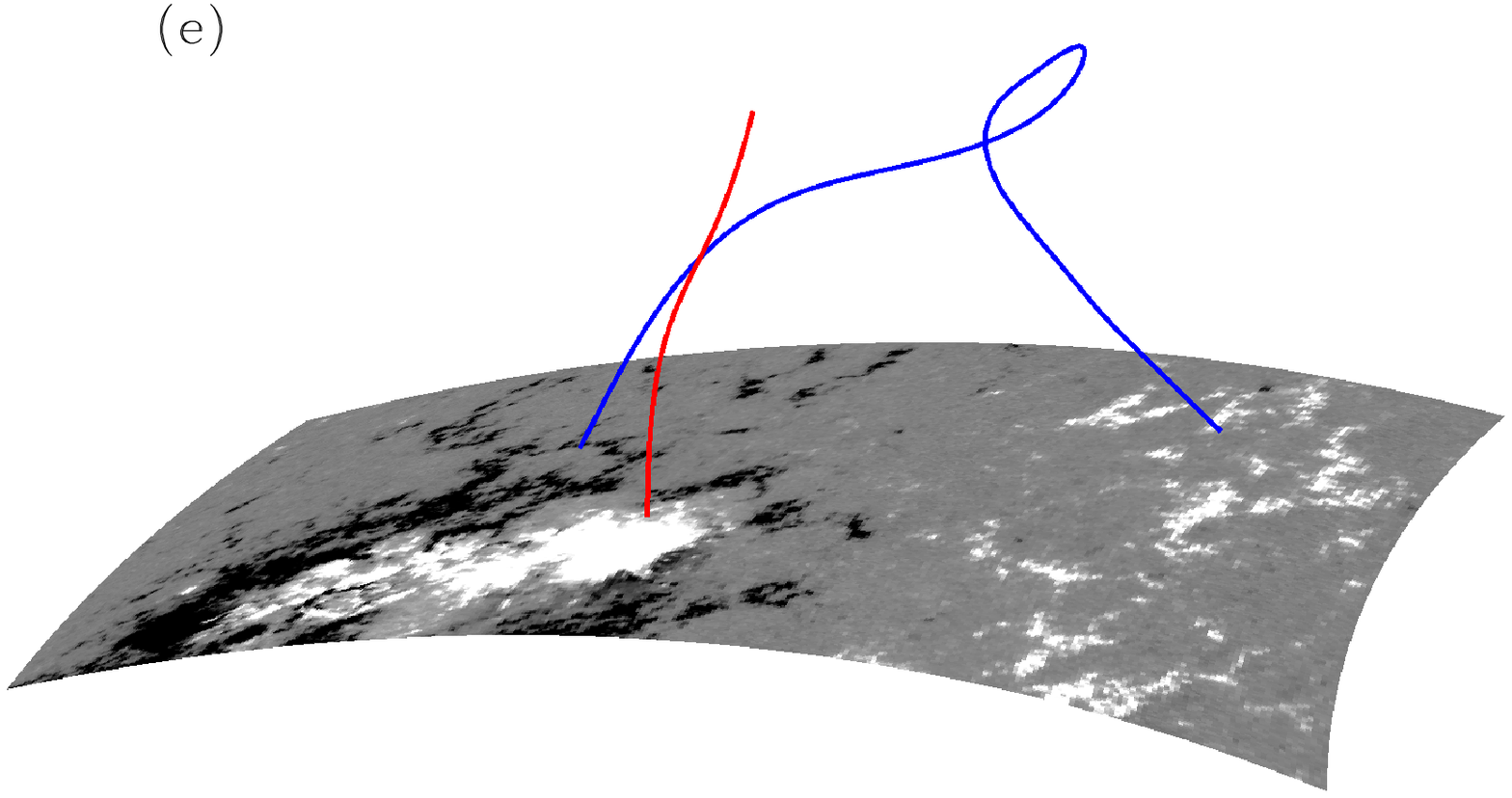}
      \includegraphics[bb =  90 441 487 627, clip, width=0.46\textwidth]{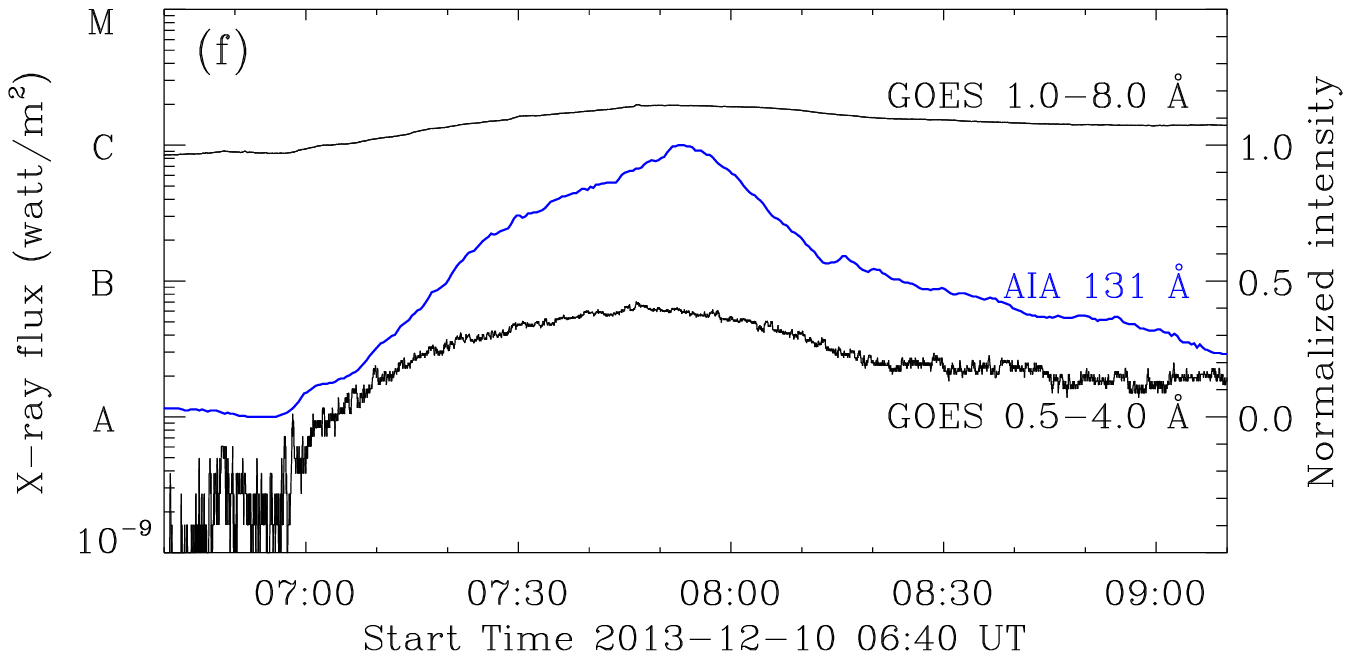}
    
    \caption[fig1]{A solar flare and associated structures observed from \Hinode, \SDO, and \STA. A horizontal slit in (b) is selected to study the outward motions.  A vertical slit in (c) and another 5-pixels-width slit in  (d) are chosen to study the inward motions for \SDO/AIA and \STA/EUVI, respectively. Profiles of two short loops in the south at 06:58~UT are marked with dotted lines.  The inset of (d) displays the relative locations of the spacecraft. Two typical converging groups of long loops are marked by the dashed lines in (c) and (d).  Their 3D reconstruction is shown in (e). The grey-scaled surface displays the differentially-rotated photospheric magnetic fields observed two days before this flare. This approach is chosen to reduce the projection effects due to the location of this active region near the west limb. (f) The light curves of \GOES ~and normalized intensity of AIA 131 \AA.\\
    ( Animations of this figure are available, see Animations 1 and 2.)}
    \label{fig:fig1}
  \end{center}
\end{figure*}

The flare under study occurred on 2013 December 10 in AR 11916 at location of W60S15. We report the observations provided by three spacecraft from two different viewing angles (Figure~\ref{fig:fig1}(d)), \ie the {\it Solar Dynamics Observatory} (\SDO; \citealp{Pesnell2012}) and \Hinode ~(\citealp{Kosugi2007})  near the Earth and the {\it Solar Terrestrial Relations Observatory} (STEREO; \citealp{Kaiser2008}) {\it Ahead} spacecraft (\STA), with a separation angle of around 150 degrees. The Atmospheric Imaging Assembly (AIA; \citealp{Lemen2012}) aboard \SDO  ~takes full-disk images of the Sun in 10 EUV/UV channels ($\log (T)$ ranges 3.7 -- 7.3), with roughly 12-second cadence. Full-disk magnetograms are provided by the Helioseismic and Magnetic Imager (HMI; \citealp{Schou2012}) on board SDO, with 1$''$ spatial resolution and 45-second cadence. The X-ray Telescope (XRT; \citealp{Golub2007}) on board \Hinode ~provides complementary observations of this active region in multiple bandpasses with a scale of $\sim$1$''$ pixel$^{-1}$. This flare appears at the eastern limb from the perspective of \STA. The Extreme-Ultraviolet Imager (EUVI; \citealp{wuelser2004euvi}) on board {\it STEREO} observed the Sun in four bandpasses, with a scale of 1.6$''$ pixel$^{-1}$ and a cadence of 5 minutes in the 195~\AA~filter.

\subsection{Results}

The X-ray peak of the flare was recorded at $\sim$07:47~UT by the {\it Geostationary Operational Environmental Satellite} (\GOES) with a magnitude of C2.0. Figure~\ref{fig:fig1}(f) shows both the evolution of the \GOES ~X-ray emission and the associated light curve in AIA 131 \AA ~($\sim$10 MK) covering this flare region. 

 \begin{figure*}[ht]
  \begin{center}
      \includegraphics[bb = 297 124 643 695, angle=90,clip, width=0.98\textwidth]{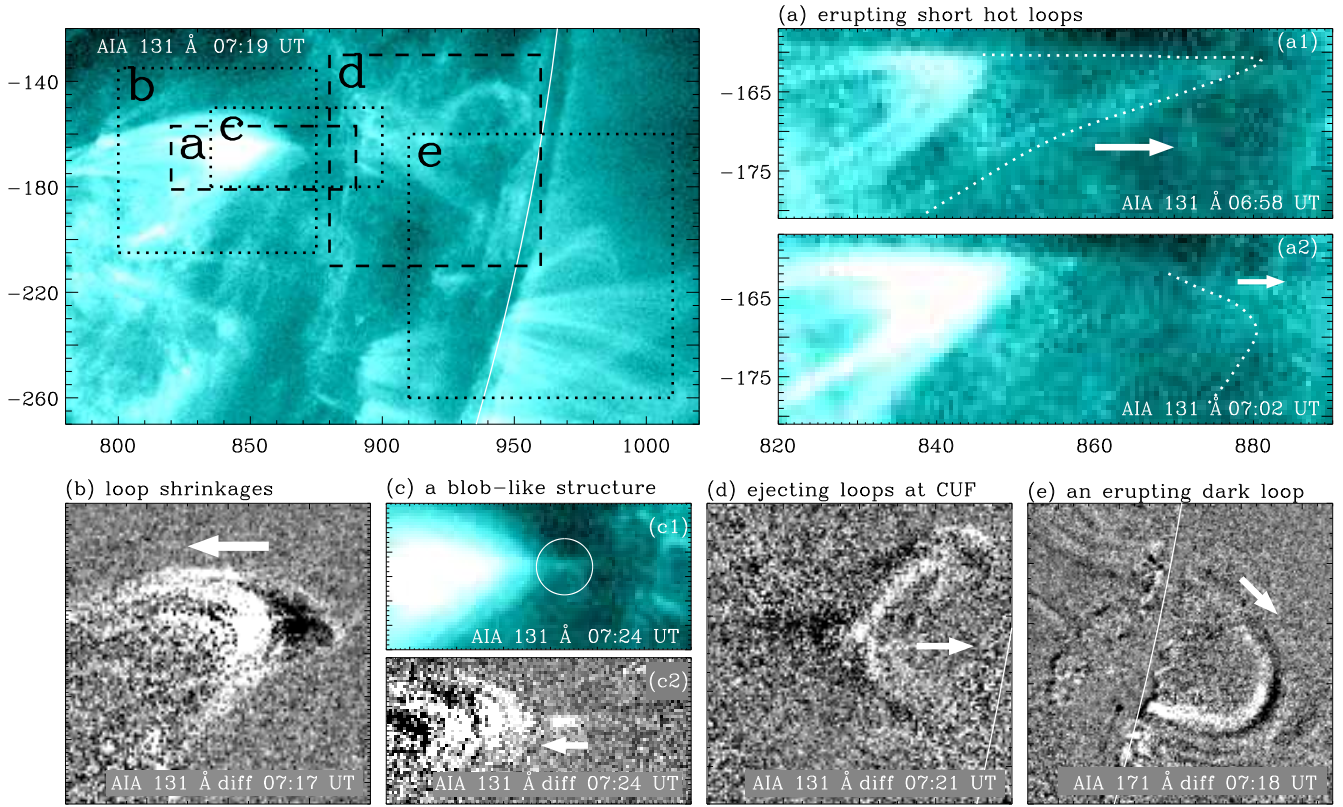}

    \caption[figf]{Several moving features observed during the solar flare. The relative positions are displayed in the top left panel. Each box is labeled to match the corresponding panels. (a) Erupting short hot loops (SHL). (b) A running difference image showing downward loop shrinkages. (c) A blob-like structure in AIA 131 \AA ~(c1) and a running difference image (c2). (d) Upward ejecting loops at the CUF. (e) A running difference image displaying an erupting dark loop beyond the solar limb. The arrow in each case indicates the direction of motion. \\
    (Animations of this figure is available, see Animations 3 and 4.)}
    \label{fig:figf}
  \end{center}
\end{figure*}

Starting at $\sim$06:54 UT, several short hot loops (SHL) (Figure~\ref{fig:figf}(a)) visible in AIA 131~\AA, erupted westward (or upward) and away from the flare loop-top region. A few of them appear as sharply-angled structures (Figure~\ref{fig:figf}(a1)) probably due to projection of the tilted loops. Meanwhile, the nearby loops observed in 193 \AA ~($\sim$1.6 MK) immediately to the north and south of this region converged, see Figure~\ref{fig:fig1}(c) and  Animation~1. The loops observed in the north were noticeably kinked, see Figures~\ref{fig:fig1}(b) and (c). In the south, both short and long loops are observed, marked in Figure~\ref{fig:fig1}(c). The 3D configurations of two groups of converging long loops (dashed lines in Figures~\ref{fig:fig1}(c) and (d)) are reconstructed using the SolarSoft routine \textsf{scc\_measure} (\citealp{thompson20123d}), see Figure~\ref{fig:fig1}(e) and Animation~2. The northern kinked loop had its eastern and western footpoints connected to the dispersed negative and positive polarities on the solar surface, respectively. The southern long loop had one root in the condensed positive polarity and deviated southward (the other part of the southern loop is not included here due to its strong background emission). These configurations suggest that these loops are not coplanar.

At $\sim$07:08~UT, a ``V-shaped'' concave-up feature (CUF) appeared in both XRT X-rays (``Thin Al\_poly'' filter, $\sim$2 -- 10 MK, see Figures~\ref{fig:fig1}(a)) and AIA 131~\AA ~(see Figures~\ref{fig:fig1}(b)). The development of this feature is clearly shown in Animation 1. Similar features can be found in the previous observations of loop-loop interaction (\citealp{su2013imaging}) and plasmoid ejections (\citealp{liu2013plasmoid}). The CUF was connected to the flare loop-top region by a thin layer. The location of this thin layer, and its appearance in the hot channels, suggest it is a CSLS.

An erupting dark loop, best identified through a series of running difference images (Figure~\ref{fig:figf}(e)), was detected at the west solar limb at 07:11~UT and appeared as a relatively dim structure in AIA 171~\AA ~($\sim$0.6 MK, see Animation~3). Its location and velocity ($\sim$60 km~s$^{-1}$) indicate this dark loop might correspond to an SHL that reached the solar limb. No obvious CME is evident  in {\it SOHO}/LASCO data   to be associated with the erupting loop structures, possibly because they were too faint at higher altitudes.

  \begin{figure*}[ht]
  \begin{center}
      \includegraphics[bb = 60 395 1114 1035, clip, width=0.98\textwidth]{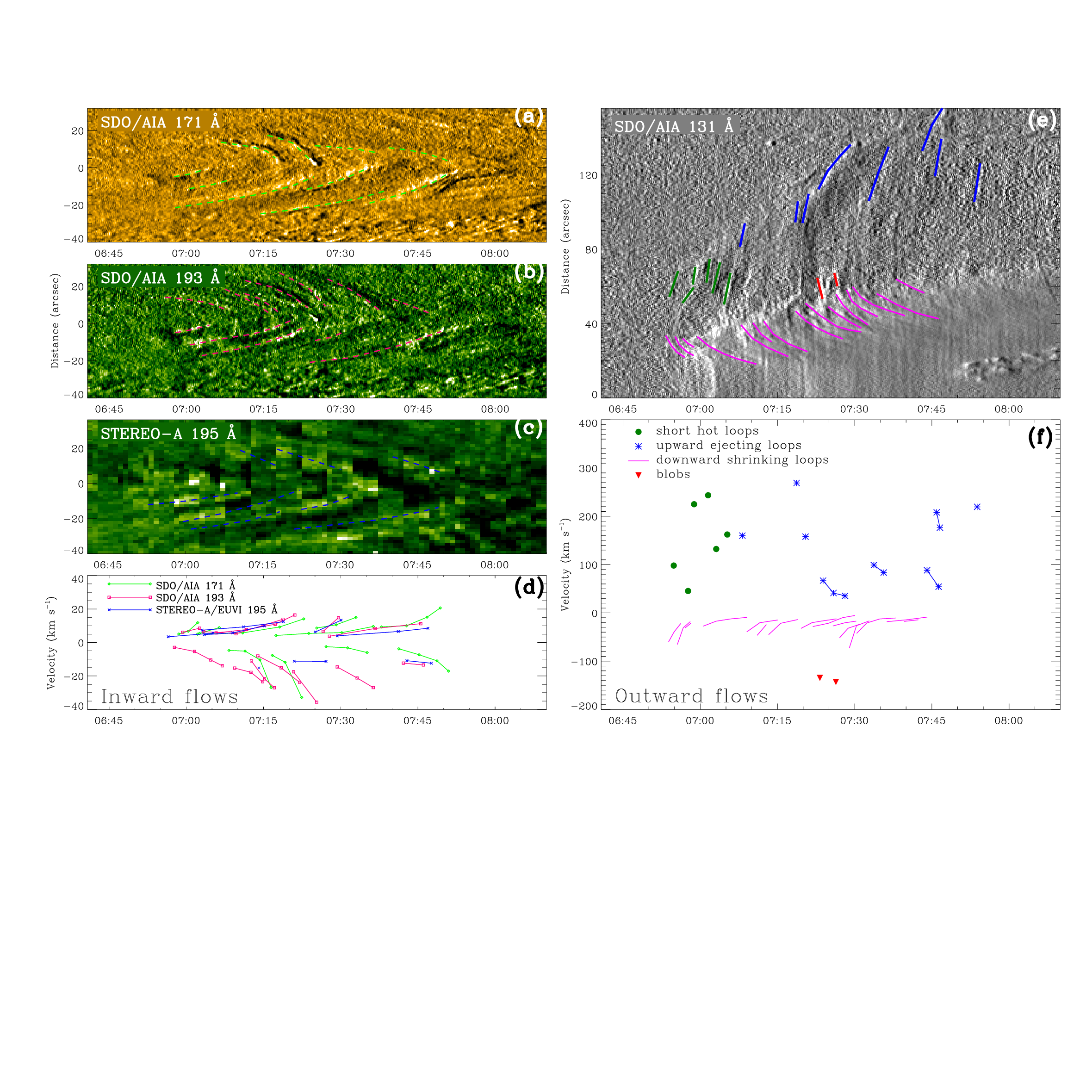}
    \caption[fig2]{Inward and outward motions. The converging flows along the vertical slit in Figure~\ref{fig:fig1}(c) are displayed in a running ratio difference of the images, with (a) from \SDO/AIA 171~\AA, (b) from \SDO/AIA 193~\AA, and (c) from \STA ~195~\AA, denoted by the dashed curves. Velocities are shown in (d). (e) A stackplot along the horizontal slit marked in Figure~\ref{fig:fig1}(b). The green lines denote the erupting short hot loops,  the red lines for the blobs. The pink lines mark the downward shrinking loops. The blue lines mark the upward ejecting loops at the CUF.  (f) Velocities of the outward motions identified in (e).\\
    (Animations of this figure is available, see Animations 3 and 4.)}
    \label{fig:fig2}
  \end{center}
\end{figure*}

A north-south virtual slit in Figure~\ref{fig:fig1}(c) is selected to generate a space-time stackplot (Figures~\ref{fig:fig2}(a) and (b)).  Several inward flows are indicated as converging streaks, with velocities derived  by linear fittings and shown in Figure~\ref{fig:fig2}(d). Inward velocities have an average value of $\sim$10 km~s$^{-1}$, and tend to increase as they approached the CSLS, similar to the observation by \citet{su2013imaging}. Two groups of loops in the north observed between 07:10 and 07:25 UT show relatively larger final velocities of $\sim$30 km~s$^{-1}$. Similar trajectories are also observed from \STA ~in EUVI 195 \AA, see Figure~\ref{fig:fig2}(c), generated along a slit marked in Figure~\ref{fig:fig1}(d).

Motions along the CSLS are studied with an east -- west oriented slit, see Figure~\ref{fig:fig1}(b). The trajectories of four moving features are presented in Figure~\ref{fig:fig2}(e), including the erupting SHL, shrinking loops at the flare apex, upward ejecting loops at the CUF, and blob-like structures (Figure~\ref{fig:figf}). 1) A series of erupting SHL (Figure~\ref{fig:figf}(a)) primarily observed in the early stage of the flare from 06:54 UT to 07:12 UT, exhibit velocities ranging from 45 to 240 km~s$^{-1}$ (Figure~\ref{fig:fig2}(f)), with an average value of $\sim$150 km~s$^{-1}$. 2) The downward shrinking loops (Figure~\ref{fig:figf}(b)), which initiated simultaneously with the erupting SHL, retract continuously towards the loop-top region until $\sim$08:10~UT. The velocities of these shrinking loops changed from between $\sim$40 -- 80 km~s$^{-1}$ to less than 10 km~s$^{-1}$ in $\sim$10 minutes. 3) Several upward ejecting loops appear successively (one example is shown in Figure~\ref{fig:figf}(d)) along with the presence of the CUF, with average initial velocities of $\sim$160 km~s$^{-1}$, significantly faster than the downward ones. This phenomenon is consistent with previous observations by \citet{liu2013plasmoid} and is thought to be related to higher density near the flare top, which can strongly decelerate downward flows. 4) Between 07:22 and 07:28~UT, two blob-like structures (with the first one shown in Figure~\ref{fig:figf}(c)) moved downward with velocities of approximately 135 and 143 km~s$^{-1}$, respectively. These values are typically found in blobs (see \citet{shen2011numerical} and references therein).

 \begin{figure*}[ht]
  \begin{center}
      \includegraphics[bb = 199 98 600 710, angle=90, clip, width=0.98\textwidth] {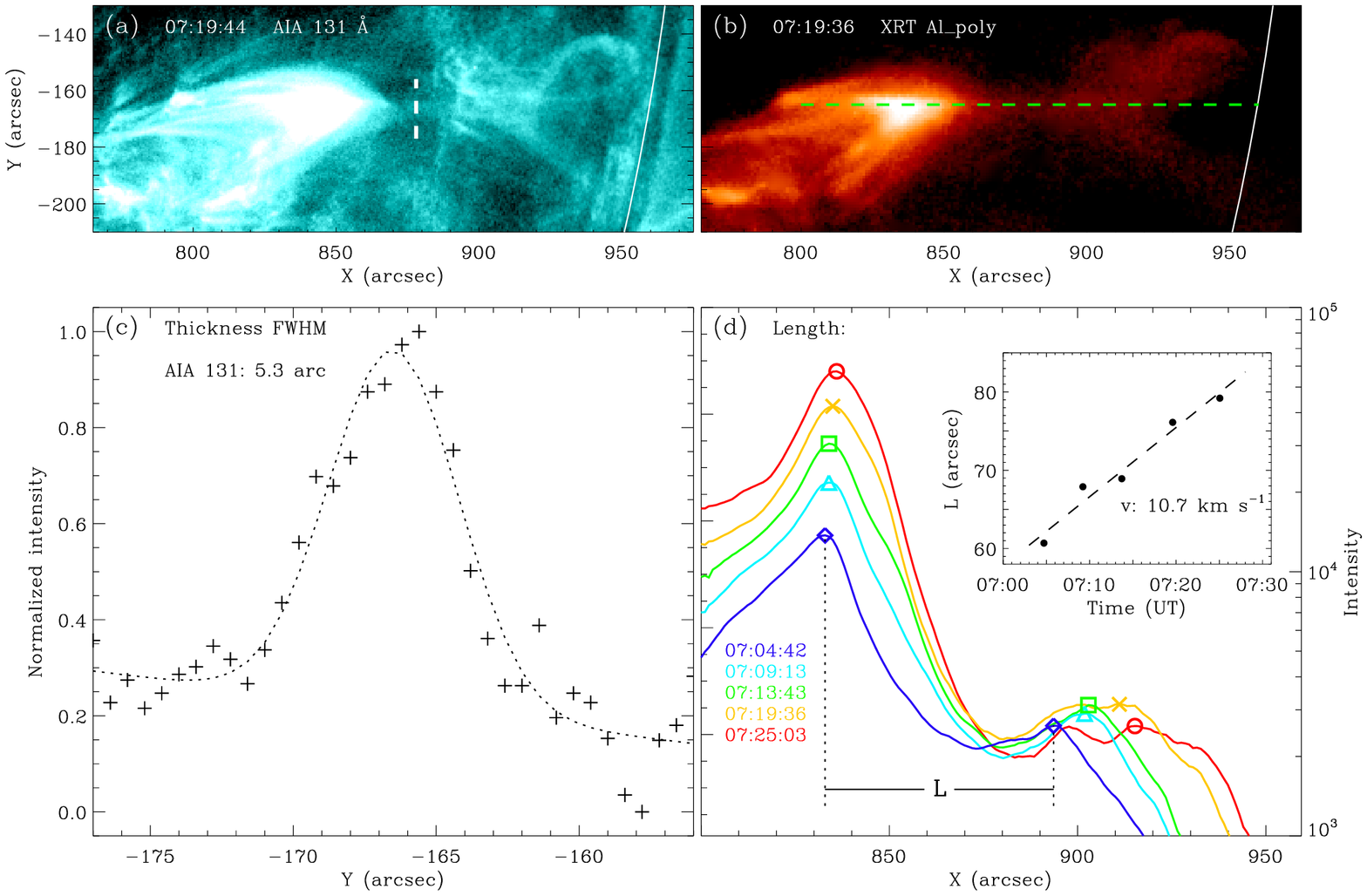}
      \includegraphics[bb = 90 210 1024 423, clip, width=0.94\textwidth]{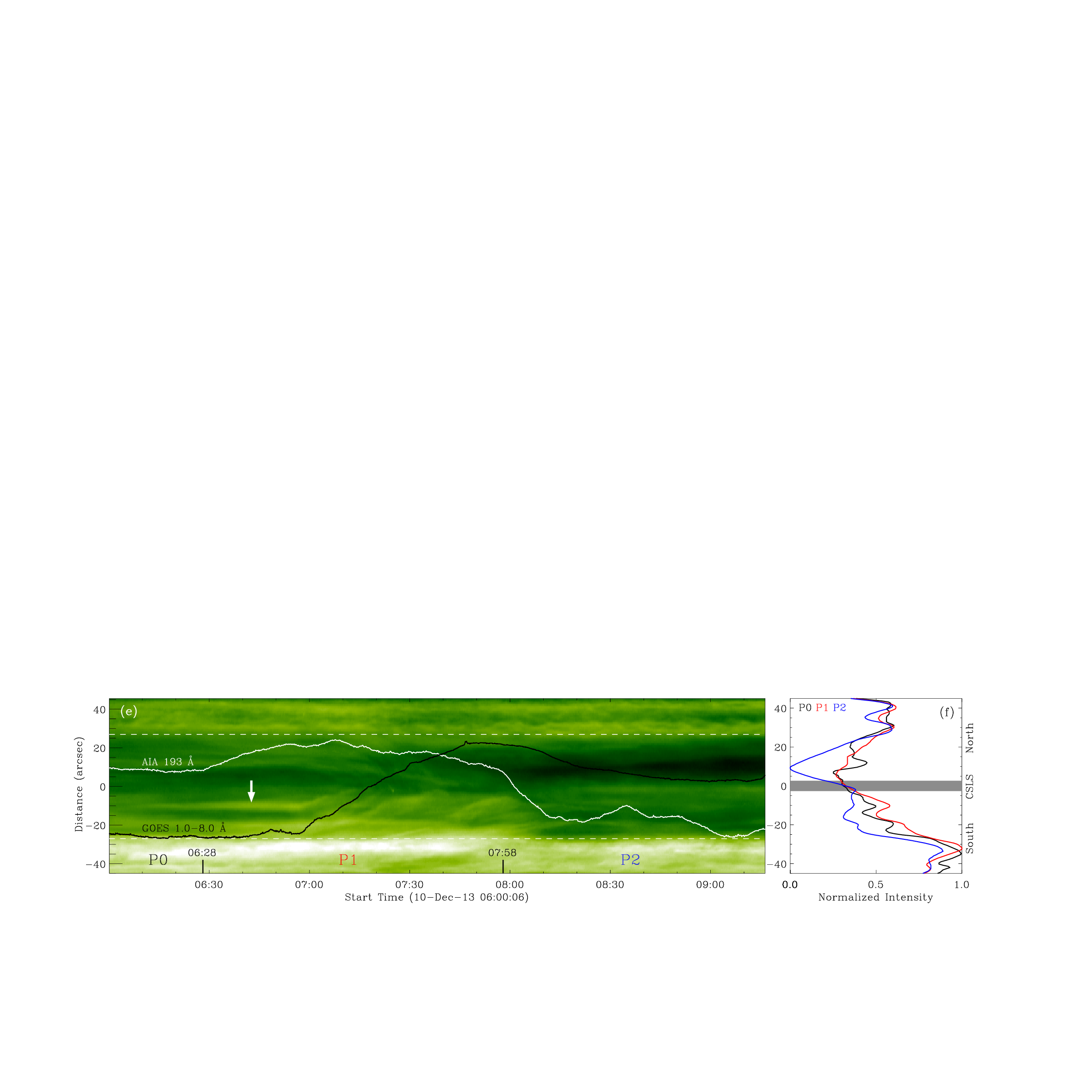}
    \caption[figcs]{The thickness and length of the CSLS, and the nearby variation of the emission. (a) AIA 131 \AA~ image at 07:19 UT. (b) XRT Al\_poly filtered image at the same time. (c) Normalized intensity along the vertical slit  in (a), with Gaussian fit marked by the dotted curve. The Full width at half maximum (FWHM) of the Gaussian fit is 5.3 arcsec. (d) Intensities along the horizontal slit marked in (b). The length of the CSLS, denoted by the distance between two localized maxima (marked by the same symbols) along each smoothed curve at five times, indicating an ongoing extension of the CSLS. A linear fitting to the growth in length is displayed in the inset. (e) Stackplot along the vertical slit in \ref{fig:fig1}(c). The white curve marks the normalized total intensities of the vertical pixels between the two dashed lines. It displays three phases of P0 -- P2. The black curve denotes the lightcurve of \GOES ~X-rays. A region with increasing emission is denoted by the arrow. (f) Normalized average values of the horizontal pixels within the three phases. The grey region, with a width of 5.3 arcsec, denotes the CSLS.\\
    (Animations of this figure is available, see Animations 1 and 5.)}
    \label{fig:figcs}
  \end{center}
\end{figure*}

The projected dimensions of the CSLS are studied at 07:19~UT, by which time it was fully developed, see Figure~\ref{fig:figcs}. The normalized intensities along a vertical slit in AIA 131~\AA ~(Figures~\ref{fig:figcs}(a)) are shown in Figure~\ref{fig:figcs}(c). A Gaussian fitting applied to this profile gives a full-width-at-half-maximum (FWHM) value of 5.3 arcsec (3.8 Mm).  The apparent thickness is necessarily an upper limit because the presence of several factors: 1) projection effects on the width of the current sheet (\citealp{lin2009investigation}), 2) a possible ``thermal halo"  around it (\eg \citealp{Yokoyama2001}), 3) the 3D shape of the current sheet (\citealp{fan2007onset}), and 4) projection of nearby hot loops onto the CSLS region. However, the value attained agrees well with past observations, such as 5 -- 10 Mm in \citet{liu2010reconnecting}, and 4 -- 5 Mm in \citet{savage2010reconnection}.

The intensities of the X-rays along the CSLS display two localized maxima: a strong source at the flare loop top, and a much weaker source at the CUF, see Figure~\ref{fig:figcs}(d). The length of the CSLS is estimated by measuring the distance between the two X-ray maxima. The evolution of this length is displayed in the inset of Figure~\ref{fig:figcs}(d). The CSLS extended from 60.7 arcsec (44 Mm) at 07:05~UT to 79.2 arcsec (57 Mm) at 07:25~UT, increasing with a velocity of 10.7 km~s$^{-1}$ given by a linear fit, and indicated by a dashed line.  As the flare progresses, the thin layer rises and extends until it is barely detectable after 07:50 UT. With the estimated length $L$ $\sim$50 Mm, and thickness $d$ $\sim$4 Mm, the measured area of the CSLS, $A_\text{cs}$, is $\sim$200 {Mm}$^2$. 

The intensity of emission in AIA 193 \AA ~around the CSLS varies as the flare progresses. A stackplot along the north-south slit (Figure~\ref{fig:fig1}(c)) between 06:00 -- 09:18 UT  is displayed in Figure~\ref{fig:figcs}(e). The white curve (used as ``Lightcurve 193'' in the following) denotes the normalized total intensities of the vertical pixels within $\pm$27 arcsec (dashed lines) neighboring the CSLS. It displays three phases, denoted by P0 -- P2, respectively. 1) P0 (06:00 -- 06:28 UT): Lightcurve 193, and that from \GOES ~(marked by a black curve), are almost flat during this phase, providing a background level in this region. 2) P1 (06:28 -- 07:58 UT): The intensity is larger than the background. It is interesting to note that Lightcurve 193 begins to increase more than 10 minutes earlier than that of \GOES. Without noticeable changes in the background (see Animation~5), the increased emissions were probably caused by two factors: i) brightenings of the loops within this region, such as those marked by an arrow in  Figure~\ref{fig:figcs}(e), that become more evident in Phase P1; and ii) brighter material/loops moving into this region. Lightcurve 193 peaks during Phase 1 at $\sim$07:10 UT and then decreases. 3) P3 (07:58 -- 09:18 UT):  Lightcurve 193 falls below the levels of P0 and P1, indicating reduced emission.  A relatively steep reduction appears between 07:58 -- 08:10 UT, during the decay phase of the flare. The lower emission might be related to coronal dimming, as reported by previous observations (\eg \citealp{sterling1997yohkoh,zarro1999soho} and see references therein) and the simulation by \citet{reeves2010current}.

A comparison of the normalized average intensities of the horizontal pixels during each of the phases, P0 -- P2, is shown in Figure~\ref{fig:figcs}(f). 1) Within 27 arcsec south of the CSLS, it is clear that the intensities in P1$>$P0. At the same time, and within the same range of the northern portion of the flaring region, P1 has lower intensity than P0 near the CSLS and higher intensity than P0 further from the CSLS. These variations suggest that the increased intensity  during P1 originated primarily from the southern side. 2) The intensities on both sides of the CSLS during P2, compared with P0 and P1, significantly decreased. This reduction of the EUV emission adjacent to the CSLS is consistent with the numerical study of coronal dimming (\citealp{reeves2010current}). 3) The southern side of the CSLS tends to be brighter than the north, probably due to the line-of-sight integration effect of the emissions from distinct coronal loops, or different temperatures and/or densities of the loops neighboring the CSLS.  

 \begin{figure}[ht]
  \begin{center}
      \includegraphics[bb = 165 393 667 710, angle=90,clip, width=0.5\textwidth]{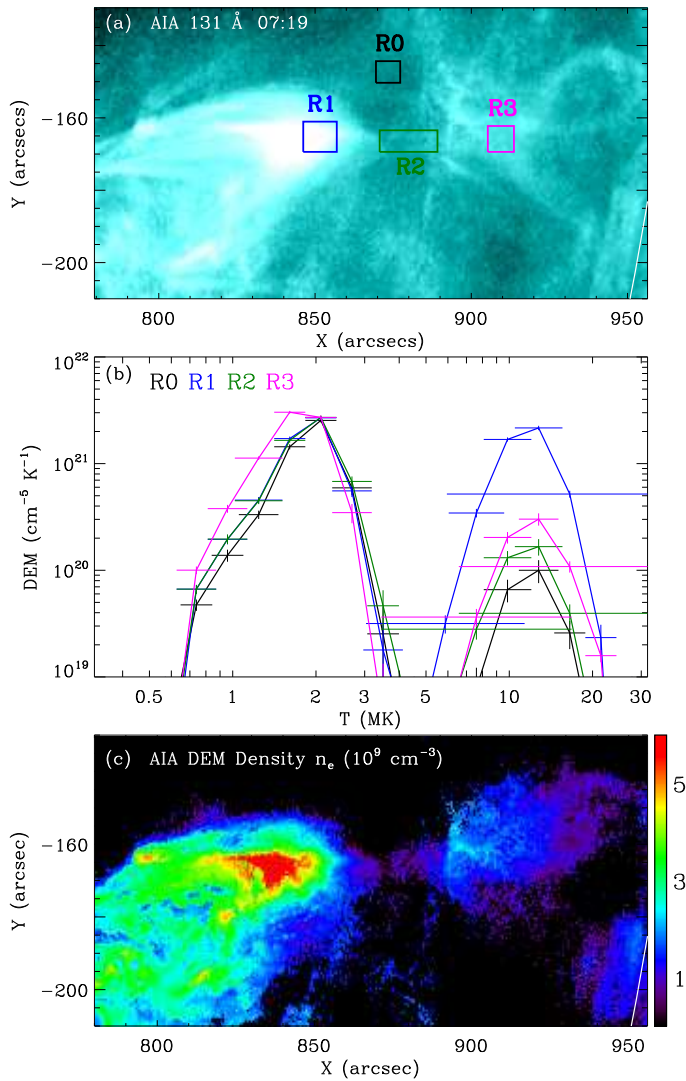}

    \caption[fig3]{(a) Four regions, R0 -- R3, are chosen to study DEM profiles, as displayed in (b). (c): Density map estimated from the DEM method with \SDO/AIA images, at 07:19 UT.}
    \label{fig:fig3}
  \end{center}
\end{figure}
 
The plasma properties of the structures are studied with a regularization method using AIA images to determine differential emission measures (DEMs; \citealp{hannah2012differential}). The DEM profiles are investigated in three brightening regions (R1 -- R3) associated with the flaring structures and a neighboring reference box R0, see Figure~\ref{fig:fig3}(a). They all display a bimodal distribution: a cold component peaking at $\sim$2 MK, and a hot component ranging from 7 -- 16 MK. The cold components from the flaring regions R1 -- R3 have similar DEM values to the reference region R0, suggesting this component probably came from the foreground and background in line with the flaring region (see \citealp{battaglia2012rhessi}). The hot components, however, varied among the regions. This suggests that the hot components are associated with the flare and that they have temperatures of 7 -- 16 MK. The appearance of the hot component at R0 is likely an artifact due to the SDO/AIA response calibration (\eg \citealp{battaglia2012rhessi}) or may be related to the apparently prevalent hot plasma in the solar corona (\citealp{schmelz2009some}). Another issue with the regularization method of AIA images is that it is limited to a temperature up to $\sim$18 MK, corresponding to the formation temperature of Fe {\scriptsize XXIV} (\citealp{ODwyer2010}). Thus, the presence of any hotter plasma would not be detected.

By assuming that the depth of the CSLS, $D$, is comparable to ${A_\text{cs}}^{1/2}$, and that the plasma is uniformly distributed along its depth, a lower limit to the electron density can be estimated with $n_e=\sqrt{\text{EM}/D}$, where EM is the emission measure.  The derived density map is shown in Figure~\ref{fig:fig3}(c). The estimated densities are 3.0 -- 6.8$\times 10^9$ cm$^{-3}$ at the flare top, $\sim$0.8$\times 10^9$ cm$^{-3}$ at the CSLS, and $\sim$1.6$\times 10^9$ cm$^{-3}$ at the CUF.

The magnetic reconnection rate, as denoted by the inflow Alfv\'en Mach number $M_\text{A}=V_\text{in}/V_\text{A} \approx V_\text{in}/{V_\text{out}}$, is evaluated in two ways. 1) By using the observed velocities of the inward ($V_\text{in}$ $\sim$10 km~s$^{-1}$) and outward ($V_\text{out}$ $\sim$200 km~s$^{-1}$) flows,  yielding $M_\text{A} \approx 0.05$. 2) Using the temperature of the flaring regions ranging from 7 -- 16 MK and the assumption that the magnetic field energy is equally transformed into kinetic and thermal energy (\citealp{Ko2003}), the Alfv\'en speed $V_\text{A}\approx198(\frac{T}{1~\text{MK}})^{1/2}$ km~s$^{-1}$ (\citealp{liu2011observing}) is 524 -- 792 km~s$^{-1}$. With this value of $V_\text{A}$, the reconnection rate is 0.01 -- 0.02. The relatively slower outflows, compared with the Alfv\'en speed, are probably caused by drag on these flows (\citealp{savage2011quantitative}).

\section{\uppercase{Discussion}}
In this article, we study the formation and evolution of a CSLS, associated with loop-loop interactions, by using joint multi-wavelength observations with high spatial and temporal resolutions from \SDO/AIA, \STA/EUVI and \Hinode/XRT. 

This study provides rare observational evidence and features for loop-loop interaction. 1) Several signatures of ongoing reconnection appear in a single event, such as converging motion, downward shrinking loops and the upward ejecting loops, blob-like structures, and a CSLS connecting the flare loop top and the CUF. 2) The reconstructed topology of two groups of interacting loops (Figure~\ref{fig:fig1}) indicate they are noncoplanar and involve kinked loops in the north, revealing the 3D aspects of this solar flare, and 3) The whole process involves the interactions between the kinked loops in the north with both long and short loops in the south. The appearance of the CSLS is probably a result of these interactions.

Our study reveals several important physical properties of the CSLS in loop-loop interaction. First, the length of the CSLS appeared to grow slowly with  an average speed of $\sim$11 km~s$^{-1}$. The extension of the current sheet is expected during solar flares, and can grow as fast as a few hundred km~s$^{-1}$ in the wake of an erupting CME (\citealp{forbes2000can,savage2010reconnection}). The low growth speed here might be related to a different scenario, \ie loop-loop interaction.  Second, the observed erupting SHL, which initiated along with the downward shrinking loops, and appeared intermittently with an average velocity of $\sim$150 km~s$^{-1}$, might be related to the outward flows/blobs in current sheet (\eg \citealp{shen2011numerical}). Third, the emission in AIA 193~\AA ~adjacent to the CSLS first increased, before its steep reduction in the late stage of the flare. This reduction in the emission is suggestive of the coronal dimmings frequently reported in association with CMEs (\eg \citealp{sterling1997yohkoh,zarro1999soho}).  In the present case, the dimmings are reported to result from a loop-loop interaction without evident detection of a CME.  The relationship between the earlier increase in emission  and the onset of the flare, if any, needs to be investigated in future studies.

\ \\ 
\ \\

\acknowledgments

The authors would like to thank the referees for many valuable comments that helped improve this article. We thank Wei Liu for helpful discussion. R.T.J.M. and C.Z. are funded by NSF-CAREER 1255024. D.A. acknowledges support by the NSF SHINE grant AGS-1061899. R.L. acknowledges the Thousand Young Talents Programme of China, and NSFC 41474151.  We also thank SDO, STEREO, Hinode, SOHO and GOES for the data support.

\end{document}